# What kind of noise guarantees security for the Kirchhoff-Loop-Johnson-Noise key exchange?

ROBERT MINGESZ, GERGELY VADAI and ZOLTAN GINGL

*Department of Technical Informatics, University of Szeged*
*Árpád tér 2, 6720 Szeged, Hungary*

This article is a supplement to our recent one about the analysis of the noise properties in the Kirchhoff-Law-Johnson-Noise (KLJN) secure key exchange system [Gingl and Mingesz, PLOS ONE 9 (2014) e96109, doi:10.1371/journal.pone.0096109]. Here we use purely mathematical statistical derivations to prove that only normal distribution with special scaling can guarantee security. Our results are in agreement with earlier physical assumptions [Kish, Phys. Lett. A 352 (2006) 178-182, doi: 10.1016/j.physleta.2005.11.062]. Furthermore, we have carried out numerical simulations to show that the communication is clearly unsecure for improper selection of the noise properties. Protection against attacks using time and correlation analysis is not considered in this paper.

*Keywords:* KLJN; secure key exchange; unconditionally secure communication; secure key distribution; noise

# INTRODUCTION

At present the security of the communication is mostly provided by software-based cryptographic solutions. Since the security is ensured only by the assumption that the eavesdropper does not have enough processing capability to break the code, considerable efforts have been made to develop unconditionally secure communication protocols. One promising research area is the quantum encryption, where security is based on the laws of quantum mechanics. However, recently an alternative communication scheme has been proposed, the Kirchhoff-Law-Johnson-Noise (KLJN) protocol, which is based only on the laws of classical physics [2]. One of the he main advantages of the KLJN protocol is that it can provide at least the same security as quantum systems at orders of magnitude lower cost. Although until now there are only a few real implementations of the system [3,4], many potential applications, such as key distribution over Smart Grid [5], uncloneable hardware keys [6] or securing computer hardware [7] have been proposed. While several attack methods have been discussed [8-13], the debate is still going on concerning the security of the system [14, 15].

The simplified diagram of the communication system is shown on Fig. 1. During the key exchange both Alice and Bob randomly select a L or H bit value.. Then, they select the corresponding resistor ($R_L$ and $R_H$) and connect it to the wire. The noise sources, $V_L(t)$ and $V_H(t)$ represent the thermal noise of the resistors. During the communication, the voltage and current noise measured in the wire ($V_E(t)$ and $I_E(t)$) are determined by the selected resistors and can be measured not only by Alice and Bob, but also by the eavesdropper, Eve. The security of the system is based on the assumption that even if Eve can measure these signals, she cannot differentiate between the LH state and HL state.

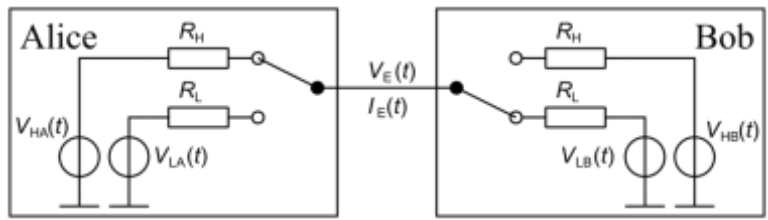

Fig. 1. Simplified diagram of the KLJN system (HL state is shown)

In real applications the thermal noise of resistors is too low; therefore, voltage noise generators are typically used to emulate high enough temperature [13]. It has already been stated that the security requires the use of Johnson-like noise, namely the noise must have normal distribution and the standard deviance must be scaled as the root of the resistance [2]. We have proven this statement using purely mathematical statistical tools [1], and in the present article we will show that these noise properties not only needed, but also guarantee absolute security against statistical attacks. Note that in this paper we do not address protection against attacks based on the analysis of the time dependence of the signals.

# RESULTS

Eq. (1) and Eq. (2) show the voltage and current values that can be measured by the eavesdropper during the two secure states, LH and HL, respectively. The notation used in the equations is introduced in Fig. 1. The communication is secure if Eve cannot distinguish between these two states. The voltage and current measured by Eve in the LH state:

$$V_{E,LH}(t) = \frac{V_{LA}(t) \cdot R_H + V_{HB}(t) \cdot R_L}{R_L + R_H} \text{ and } I_{E,LH}(t) = \frac{V_{HB}(t) - V_{LA}(t)}{R_L + R_H}. \quad (1)$$

The voltage and current signal measured by Eve in the HL state:

$$V_{E,HL}(t) = \frac{V_{HA}(t) \cdot R_L + V_{LB}(t) \cdot R_H}{R_L + R_H} \text{ and } I_{E,HL}(t) = \frac{V_{LB}(t) - V_{HA}(t)}{R_L + R_H}. \quad (2)$$

For secure communication, the joint probability density function $p_{LH}(I_E, V_E)$ and $p_{HL}(I_E, V_E)$ must be the same. If $I_E$ and $V_E$ are independent, this is satisfied.

As it has been proven [16], linear combinations $Y_A$ and $Y_B$ of two independent random variables $X_1$ and $X_2$ in Eq. (3) will be statistically independent if and only if each random variable is normally distributed and Eq. (4) is satisfied:

$$Y_A = A_1 \cdot X_1 + A_2 \cdot X_2 \text{ and } Y_B = B_1 \cdot X_1 + B_2 \cdot X_2, \quad (3)$$

$$A_1 \cdot B_1 \cdot \sigma_1^2 + A_2 \cdot B_2 \cdot \sigma_2^2 = 0, \quad (4)$$

where $\sigma_1$ and $\sigma_2$ are the standard variation of $X_1$ and $X_1$ respectively. In our case we obtain:

$$V_{E,HL}(t) = V_{HA}(t) \cdot R_L + V_{LB}(t) \cdot R_H \text{ and } I_{E,HL}(t) = V_{LB}(t) - V_{HA}(t), \quad (5)$$

$$R_L \cdot \sigma_{VH}^2 - R_H \cdot \sigma_{VL}^2 = 0, \quad (6)$$

where $\sigma_{VH}$ and $\sigma_{VL}$ are the standard deviations of $V_{HA}$ and $V_{LB}$, respectively. Note, that we get a similar equation for the LH case. According to this, the distribution of $V_{HA}$ and $V_{LB}$ must be normal, and the scaling of the standard deviation must follow the rule:

$$\sqrt{\frac{R_L}{R_H}} = \frac{\sigma_{VL}}{\sigma_{VH}}, \tag{7}$$

in agreement with the results presented in [1]. We have carried out numerical simulations to obtain the joint statistics of $I_E$ and $V_E$. We have generated $2^{13}$ samples both for the current and voltage and made scatter plots for several cases. Figure 2 demonstrates what happens with the joint distribution of $I_E$ and $V_E$ if Eq. (7) is not satisfied: there is an asymmetry in the distribution that depends on the actual state, LH or HL.

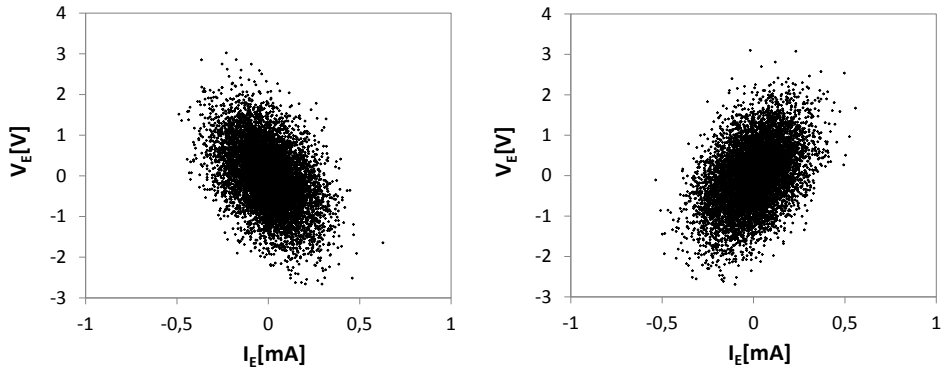

Fig. 2. Scatter plot for cases LH (left) and HL(right) using noise with normal distribution if the Eq. (6) is not satisfied. $R_L$=1 kΩ, $R_H$=10 kΩ, $\sigma_{VH}/\sigma_{VL}$=1,5.

In order to achieve a secure communication, the linear combination of noises must give the same type of probability distribution as the original one [1]. Such distributions are called stable distributions, here we consider symmetric α-stable distributions that include normal distribution as a special case. Assuming distributions symmetric around zero their characteristic function is defined by the following equation:

$$\varphi(t) = \exp\left(-\left|\frac{t}{w}\right|^\alpha\right), \tag{8}$$

where $\alpha$ is the stability parameter in the range from 0 to 2 and $w$ is the scaling factor of the probability density function. Note that $\alpha = 2$ corresponds to normal distribution and $\alpha = 1$ corresponds to Cauchy distribution. However, according to [16], $I_E$ and $V_E$ are not independent except in the case of normal distribution ($\alpha = 2$) as can be seen on Fig. 3. Note that not all of such distributions have finite variance, therefore the scaling of the noise voltages was based on the scaling factor *w* which can be associated with the voltage noise magnitude and is defined in Eq. (8). Thus, the higher and lower noise have scaling factors $w_{VH}$ and $w_{VL}$, respectively.

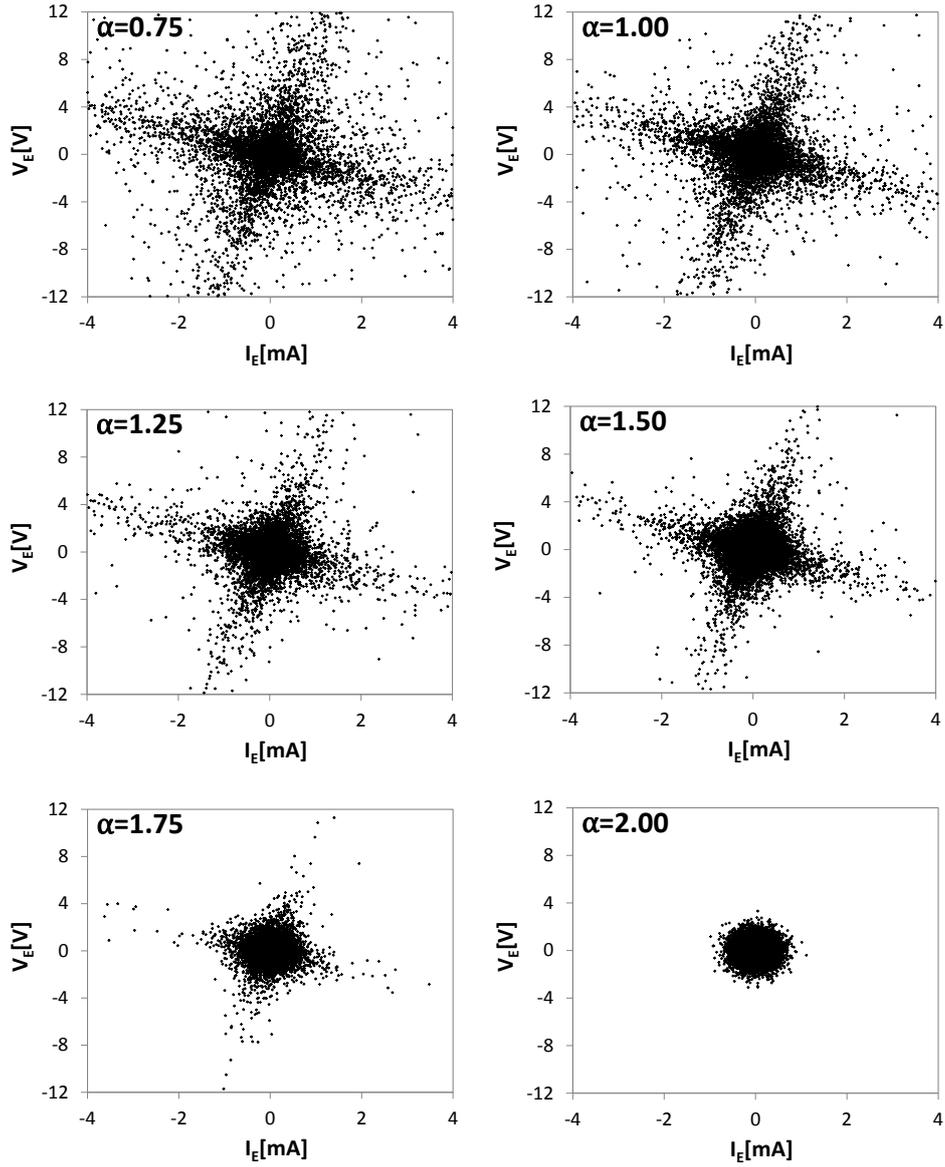

Fig. 3. Scatter plot for case HL using distributions with different values of α. Note that α = 1 and α = 2 correspond to Cauchy and normal distribution, respectively. $R_L$=1 kΩ, $R_H$=10 kΩ, $w_{VH}/w_{VL} = \sqrt{R_H/R_L}$.

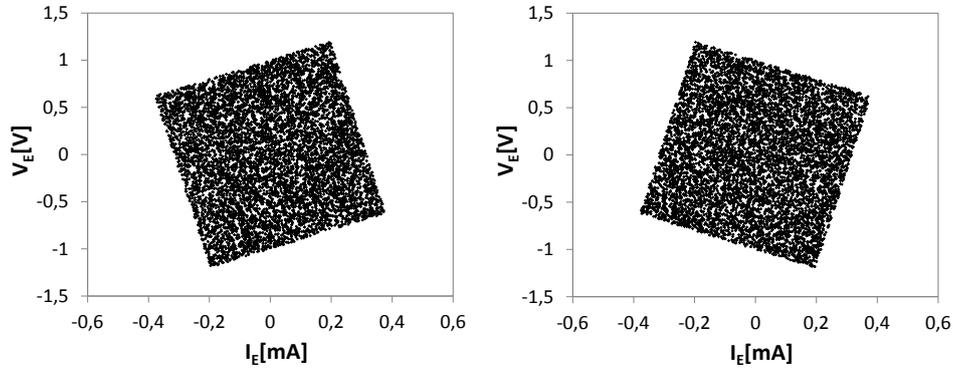

Fig. 4 Scatter plot for cases LH (left) and HL (right) using noise with uniform distribution. $R_L$=1 kΩ, $R_H$=10 kΩ, $\sigma_{VH}/\sigma_{VL} = \sqrt{R_H/R_L}$.

# CONCLUSION

We have shown that communication using the KLJN protocol is secure if and only if noise voltages with normal distribution are used and the variance of the noise voltages follow the scaling defined by Eq. (7). This result is based on mathematical statistical derivation and it is in agreement with previous results [1,2]. Note that protection against attacks using time and correlation analysis is not considered and can be addressed in subsequent publications. Further analysis can clarify how the time domain properties of the noise influence the security of the system.

# ACKNOWLEDGEMENTS

We thank the anonymous reviewer of our previous paper [1] for drawing our attention to the problem of statistical dependence of voltage and current fluctuations in the communication wire related to absolute security.

This research was supported by the European Union and the State of Hungary, co-financed by the European Social Fund in the framework of TÁMOP 4.2.4. A/2-11-1-2012-0001 'National Excellence Program'.